\documentclass[aps,superscriptaddress,twocolumn,twoside,floatfix,pra,nofootinbib,a4paper]{revtex4}
\usepackage{times}
\usepackage{epsfig}
\usepackage{amsfonts}
\usepackage{amsmath}
\usepackage{amssymb}
\usepackage{amsthm}
\usepackage{color}
\usepackage{multirow}
\usepackage[normalem]{ulem}
\newcommand{\stkout}[1]{\ifmmode\text{\sout{\ensuremath{#1}}}\else\sout{#1}\fi}
\usepackage{latexsym}
\usepackage{mathrsfs}
\usepackage{natbib}
\usepackage{verbatim}
\usepackage[T1]{fontenc}
\usepackage{float}

\usepackage{graphicx}
\usepackage{xcolor}

\usepackage[colorlinks=true,linkcolor=blue,citecolor=magenta,urlcolor=blue]{hyperref}

\DeclareMathOperator{\Tr}{tr}

\newcommand{\ket}[1]{|#1\rangle}

\newcommand{\braket}[2]{\langle#1|#2\rangle}
\newcommand{\bracket}[3]{\langle#1|#2|#3\rangle}
\newcommand{\ketbra}[2]{|#1\rangle\langle#2|}

\begin{document}


\title{Semi-device-independent certification of independent quantum state and measurement devices}


\author{Armin Tavakoli}
\affiliation{D\'epartement de Physique Appliqu\'ee, Universit\'e de Gen\`eve, CH-1211 Gen\`eve, Switzerland}

\begin{abstract}
Certifying that quantum devices behave as intended is crucial for quantum information science. Here, methods are developed for certification of both state preparation devices and measurement devices based on prepare-and-measure experiments with independent devices. The experimenter assumes the independence of the devices and knowledge of the Hilbert space dimension. Thus no precise characterisation of any part of the experiment is required. The certification is based on a randomised version of unambiguous state discrimination and targets the class of state ensembles corresponding to quantum $t$-designs of any size and any dimension. These quantum designs are sets of states over which the average of any $t$-degree polynomial equals its average over all pure states, and they accommodate many of the most useful discrete structures in quantum information processing. Furthermore, it is shown that the same experiments also certify the detection efficiency of the measurement devices, as well as their non-projective nature. The presented methods can readily be implemented in experiments.
\end{abstract}


\maketitle


\textit{Introduction.---} The precise control of quantum devices is crucial for the development of quantum technologies and experimental tests of foundational phenomena in quantum theory. Therefore, methods for certifying and characterising quantum devices are indispensable in quantum information science. Such methods allow one to ensure that, for example, a state preparation device indeed prepares the intended state or that a measurement device implements the desired measurements. The most common approaches are quantum state tomography \cite{Vogel} and quantum detector tomography \cite{Luis}. In the former, the state emitted by a preparation device is measured in several different bases and the resulting outcome statistics is used to determine the state. In the latter, the measurement is determined from  the outcome statistics obtained from probing it with different states. Therefore, the success of state (detector) tomography hinges on the auxiliary measurement (state preparation) device being precisely calibrated. Consequently, imperfections on experimentally relevant parameters in the auxiliary devices can undermine both state tomography \cite{Rosset} and detector tomography \cite{Anwer}. Moreover, in order to precisely calibrate the auxiliary device itself, one typically also requires a tomographic procedure which leads to an infinite regress.

The requirement of precise control in tomography can be overcome by more sophisticated certification methods. Methods have been developed for certifying states and measurements in experimental settings in which a sender prepares states and a receiver measures them  but without neither device  requiring a detailed characterisation \cite{Tavakoli18}. Instead, the only assumption is that the Hilbert space dimension is known, which can often be justified from inspecting the specific experimental setup. This so-called semi-device-independent (SDI) approach to certification of quantum devices benefits from the fact that prepare-and-measure experiments are practical to implement \cite{Hendrych, Ahrens, Ambrosio, Lunghi, TH15, Martinez} while also allowing for realistic experimental imperfections. A variety of  SDI certification methods have been developed, e.g.~for qubit states and measurements \cite{Tavakoli18}, pairs of mutually unbiased bases and symmetric informationally complete measurements \cite{Farkas19, TavakoliSIC}, non-projective qubit measurements \cite{TavakoliNonProj, Piotr} and qubit quantum instruments \cite{Mohan, Miklin}. The practical viability of SDI certification schemes has also been experimentally demonstrated \cite{TavakoliNonProj, Anwer, Smania, Foletto}. 

With a few notable exceptions focused on dimension witnessing \cite{Bowles, Vicente} and random number generation \cite{Lunghi}, previous works on dimension-bounded quantum correlations in general, and SDI certification schemes in particular, have adopted models in which the involved quantum devices can be classically correlated in a stochastic and (to the experimenter) unknown manner. Such models make the set of quantum correlations convex, thereby considerably simplifying their analysis. However, these models correspond to a paranoid setting in which devices can classically conspire against the experimenter. In experiments that do not involve  malicious parties, it is often natural to assume that separate quantum devices are independent.

In this work, we develop SDI certification methods for independent preparation and measurement devices. To this end, we present a SDI variant of the well-studied task of unambiguous state discrimination \cite{Ivanovic, Dieks, Peres}. We prove that by observing optimal correlations in this task, one can certify the collection of states produced by the preparation device. The certification has a broad scope of relevance since it targets any number of states in any dimension that form a quantum $t$-design \cite{Delsarte, Renes}. A quantum $t$-design is a set of $d$-dimensional quantum states with the  property that the average of any $t$-degree polynomial over the set equals the average taken over all pure states. These interesting and highly symmetric structures have broad applications in quantum information science. Examples include quantum tomography \cite{Scott, Hayashi}, quantum key distribution \cite{SixState, Joe}, Bell inequalities \cite{MUBSIC}, entropic uncertainty relations \cite{Ketterer} and entanglement detection \cite{Lloyd, Bae}. They also accommodate (as special cases) some of the most intensely researched and celebrated discrete Hilbert space structures such as rank-one generalised measurements, complete sets of mutually unbiased bases \cite{MUBs} and symmetric informationally complete sets of states \cite{SICs}, as well as the Platonic solids \cite{Plato}.  Moreover, we also use the same task to  show that useful properties of the measurement device can be certified. Specifically, we show that one can certify the detection efficiency of the setup in a SDI manner. This is motivated by the fact that detectors do not always succeed with detecting an incoming physical system and that this is an important consideration in many quantum information protocols. Importantly, in order to make the certification of states and measurements experimentally relevant, we develop its robustness to errors. Finally, we also exemplify the application of the scheme towards certification of non-projective measurements and show that it is substantially more robust to errors than established SDI certification schemes based on classically correlated devices.

\textit{Randomised unambiguous state discrimination.---} Our platform for certifying quantum states and measurements is inspired by the textbook task of unambiguous state discrimination (USD). In USD, a sender (Alice) randomly chooses one of two possible states, $\ket{\phi_1}$ and $\ket{\phi_2}$, and sends it to a receiver (Bob). By measuring the incoming state, Bob tries to unambiguously decide which state he has received. Thus, he must either correctly identify the state or declare that he does not know the answer (inconclusive). His success rate is
\begin{equation}\label{succ}
p_\text{usd}^{\phi_1,\phi_2}=\frac{1}{2}\left(p(1|\phi_1)+p(2|\phi_2)\right),
\end{equation}
while no errors are made, i.e.~$p(1|\phi_2)=p(2|\phi_1)=0$. Naturally, as soon as Alice's two states are not perfectly distinguishable (orthogonal), Bob cannot achieve a perfect success rate and must sometimes declare inconclusive rounds.  It is well-known \cite{Peres, Ivanovic, Dieks} that Bob's best measurement is $\{M^1,M^2,M^\perp\}$ where
\begin{align}\label{musd}
& M^1=\frac{\openone-\ketbra{\phi_{2}}{\phi_{2}}}{1+|\braket{\phi_1}{\phi_{2}}|}, & M^{2}=\frac{\openone-\ketbra{\phi_{1}}{\phi_{1}}}{1+|\braket{\phi_1}{\phi_{2}}|},
\end{align}
and $M^\perp=\openone-M^1-M^{2}$, where ``$\perp$`` denotes the inconclusive outcome. This leads to the optimal success rate 
\begin{equation}\label{ivan}
\max_M p_\text{usd}^{\phi_1,\phi_2}=1-|\braket{\phi_1}{\phi_{2}}|.
\end{equation}
In USD, the overlap of Alice's two states is assumed to be known. From this, we draw inspiration in order to construct the task of \textit{randomised USD}, in which Alice's device requires no precise characterisation but is assumed to produce states of a known Hilbert space dimension.

In randomised USD, Alice generates a random input $x\in\{1,\ldots, N\}$ and subsequently produces a $d$-dimensional state $\rho_x$. The state is sent to Bob who generates a random pair of inputs  $y\equiv (y_1,y_2)$; these can be any one of the $\binom{N}{2}$ ordered pairs ($y_1<y_2$) of two positive integers no larger than $N$. He then implements a corresponding measurement  $M_y=\{M_{b|y}\}_b$ with three possible outcomes $b\in\{1,2,\perp\}$. The random input informs him that his task is to unambiguously discriminate between Alice's two states $\{\rho_{y_1},\rho_{y_2}\}$. Thus, when $x\in \{y_1,y_2\}$, the task is to perform USD whereas otherwise the round is inconsequential. Therefore, for a given input $y$, the success rate is defined as
\begin{equation}\label{usd}
p_\text{usd}^{y}=\frac{1}{2}\left(p(1|y_1,y)+ p(2|y_2,y)\right),
\end{equation}
and the unambiguity requires $p(1|y_2,y)=p(2|y_1,y)=0$ $\forall y$, where the probabilities are given by the Born rule $p(b|x,y)=\Tr\left(\rho_xM_{b|y}\right)$. The performance in randomised USD is based on all the individual USD tasks (for each $y$). To specify the figure of merit, we  first note that the rate of inconclusive rounds is simply $1-p_\text{usd}^{y}$. We consider the moments of the rate of inconclusive events accumulated over all the individual USD tasks:
\begin{equation}\label{witness}
\mathcal{S}_{t}\equiv \sum_{y_1<y_2} \left(1-p_\text{usd}^{y}\right)^{2t},
\end{equation}
where the integer $t\geq 1$ is the order of the moments. As shall soon become clear, this figure of merit is chosen due to its connection to quantum designs and for enabling a handy technical treatment. Thus, randomised USD is parameterised by the dimension $d$, the ensemble size $N$ and the order $t$. Aiming to perform USD well for every $y$ means that Alice and Bob aim to \textit{minimise} $\mathcal{S}_{t}$. Importantly, we stress that when $N>d$, it is impossible for Alice to prepare her $N$ states so that they are all pairwise distinguishable (trivialising the task). 

\textit{Certifying quantum designs.---} We show that randomised USD allows us to certify that Alice's states form a quantum $t$-design. While we presently restrict ourselves to an ideal setting in which Bob's discrimination is unambiguous for every $y$, we will later consider the more general case in which the discrimination features errors.

We use the fact that each of Bob's measurements apply to a single USD task (that corresponding to input $y$). This allows us to write
\begin{equation}\label{QQ}
\min_{\text{quantum}} \mathcal{S}_t=\min_{\{\rho\}} \sum_{y_1<y_2} \left(1-\max_{M_y}p_\text{usd}^{y}\right)^{2t}.
\end{equation}
Leveraging the fact that the devices are independent, the minimal value of $\mathcal{S}_t$ is achieved with pure states. Therefore, in order to evaluate \eqref{QQ}, we may write $\rho_x=\ketbra{\psi_x}{\psi_x}$. Although the states are $d$-dimensional, every pair $\{\ket{\psi_{y_1}},\ket{\psi_{y_2}}\}$ can be viewed as an effective qubit embedded in the larger Hilbert space. Therefore, for every $y$, it is optimal for Bob to perform a measurement analogous to that in Eq~\eqref{musd}. Hence, for given states of Alice and input $y$, the optimal success rate is analogous to Eq~\eqref{ivan}.  A simple re-arrangement of the summation then gives
\begin{equation}\label{qq}
\min_{\text{quantum}} \mathcal{S}_t=-\frac{N}{2}+\frac{1}{2}\min_{\{\psi\}} \sum_{y_1,y_2} |\braket{\psi_{y_1}}{\psi_{y_2}}|^{2t}.
\end{equation} 
At first sight, evaluating the right-hand-side seems challenging. However, the quantity subject to the minimisation is  both  well-studied and closely linked to quantum designs; it is  commonly referred to as the ($t$'th-order) \textit{frame potential} \cite{Fickus}.

A quantum design is a set $\{\ket{\phi_i}\}$ of $N$ pure $d$-dimensional states with the property that the average of any polynomial, $g_t$ of degree $t$, taken over the set is identical to the average of the same polynomial taken over all pure $d$-dimensional states. That is,
\begin{equation}
\frac{1}{N}\sum_{i=1}^N g_t(\phi_i)=\int d\phi g_t(\phi),
\end{equation}
where $d\phi$ is the Haar measure on the space of pure quantum states of dimension $d$. The polynomial $g_t$ can be written as $g_t(\phi)=\bracket{\Phi}{G_t}{\Phi}$ where $\ket{\Phi}=\ket{\phi}^{\otimes t}$ and $G_t$  is some bounded operator in the symmetric subspace of $(\mathbb{C}^d)^{\otimes t}$. It immediately follows that a quantum $t$-design also is a $t'$-design for $t'\leq t$. How does one determine whether a set of states is a quantum design? The answer is based on the frame potential. An ensemble of $N$ $d$-dimensional states constitutes a quantum $t$-design if and only if it saturates the following lower bound on the frame potential \cite{Renes}
\begin{equation}\label{design}
V_t(\{\phi\})\equiv \sum_{j,k=1}^N |\braket{\phi_j}{\phi_k}|^{2t}\geq \frac{N^2 t!(d-1)!}{(t+d-1)!}\equiv J_{t}.
\end{equation}

With this knowledge of quantum designs in hand, we can assert that the optimal quantum implementation of randomised USD obeys
\begin{equation}\label{selftest}
\min_{\text{quantum}} \mathcal{S}_t\geq \frac{1}{2}\left(J_{t}-N\right)\equiv \mathcal{Q}_{t}.
\end{equation}
The bound $\mathcal{Q}_t$ can be saturated if and only if Alice's states form a $t$-design of dimension $d$ compsed of $N$ states. Hence, this completes the certification. Moreover, notice that \eqref{selftest} also serves as a family of device-independent dimension witnesses for independent devices.

\textit{Certification under discrimination errors.---} Since a small rate of failed discriminations (incorretly identifying the state) is to be expected in any realistic experiment, let us depart from the ideal situation and consider certification of the preparations when Bob's discrimination, for each input $y$, is subject to an error rate. We show that certification of quantum designs remains possible. 

We adopt a model in which Alice's two pre-established equiprobable states $\ket{\phi_1}$ and $\ket{\phi_2}$ are to be discriminated in such a way that the rate of incorrect announcements associated to outcome $1$ and $2$ resepctively does not exceed $\epsilon\in[0,\frac{1}{2}]$. That is, the rate of error is bounded by $q_{1}\leq \epsilon$ and  $q_{2}\leq \epsilon$ where
\begin{align}\label{boundederror}
& q_1=\frac{p(1|\phi_2)}{p(1|\phi_1)+p(1|\phi_2)}, && q_2=\frac{p(2|\phi_1)}{p(2|\phi_1)+p(2|\phi_2)}.
\end{align}
Evidently, standard USD corresponds to choosing $\epsilon=0$. Interestingly, the problem of finding the optimal success rate \eqref{succ} under the bounded error conditions has been solved \cite{Hayashi2}.  Ref~\cite{Hayashi2} found that the optimal success rate is given by
\begin{equation}\label{haya}
p_{\text{usd}}(\epsilon)=
\begin{cases}
\alpha_\epsilon\left(1-|\braket{\phi_1}{\phi_2}|\right) & \text{for} \quad \epsilon\leq \epsilon_\text{c}\\
\frac{1}{2}\left(1+\sqrt{1-|\braket{\phi_1}{\phi_2}^2|}\right) & \text{for} \quad \epsilon_\text{c}\leq \epsilon,
\end{cases}
\end{equation}
where 
\begin{equation}
\alpha_\epsilon =\frac{1-\epsilon}{\left(1-2\epsilon\right)^2}\left(1+2\sqrt{\epsilon\left(1-\epsilon\right)}\right)
\end{equation}
and $\epsilon_\text{c}=1/2\left(1-\sqrt{1-|\braket{\phi_1}{\phi_2}|^2}\right)$. 

Equipped with this, we can now certify the preparation device also when the error rate $\epsilon$ is found in the data. From the  measured probabilities, one can appropriately choose $\epsilon$. Then, based on the observed error rate, we modify the original figure of merit \eqref{witness} so that it reads
\begin{equation}\label{mod}
\mathcal{S}_t^{\epsilon}=\sum_{y_1<y_2} \left(\alpha_\epsilon-p_{\text{usd}}^y\right)^{2t}.
\end{equation}
Notice that the error-free case ($\epsilon=0$) returns the original figure of merit since $\alpha_0=1$. 
 To find the optimal value of $\mathcal{S}_t^\epsilon$, we can recycle the reasoning in the previous section. One can account for the piecewise continuous feature in Eq~\eqref{haya} by noticing that $\alpha_\epsilon$ is monotonically increasing and that  the upper expression in \eqref{haya} therefore serves as an upper bound on the lower expression when $\epsilon\geq \epsilon_\text{c}$. This can be applied to our problem for every $y$ and any $\epsilon$. Then, we arrive at the error-tolerant statement  
\begin{equation}\label{robusterror}
\min_{\text{quantum}} \mathcal{S}_t^\epsilon \geq \alpha_\epsilon^{2t} \mathcal{Q}_t,
\end{equation}
which generalises \eqref{selftest}. The inequality can be saturated if  and only if Alice's states form a $t$-design with the property that the relation $\epsilon\leq \epsilon_\text{c}$ is satisfied for all pairs of states. Hence, designs can be certified also in presence of discrimination errors. For example, a celebrated family of designs are known as symmetric informationally complete (SIC) \cite{Renes}. They correspond to $t=2$ and $N=d^2$ for any $d\geq 2$. Any such design can be certified through \eqref{robusterror} as long as $\epsilon$ remains reasonably small. Specifically the bound is tight when $\epsilon\leq \frac{1}{2}\left(1-\sqrt{\frac{d}{d+1}}\right)$. For qubits ($d=2$), the critical error becomes $\epsilon \approx 9.2\%$ which is well above experimentally achieved error rates in USD \cite{Clarke}.

Furthermore, consider a situation in which we observe an error rate $\epsilon$ but Alice's preparations nevertheless do not precisely form a design. Then, we can estimate how close they are to forming a design based on the measured value of $\mathcal{S}_t^\epsilon$. If we momentarily assume pure states a natural quantifier is the frame potential, which by arguments analogous to the above satisfies
\begin{equation}\label{robust}
V_t\leq N+\frac{2}{\alpha_\epsilon^{2t}}\mathcal{S}_t^{\epsilon}.
\end{equation}
Hence, the closer $\mathcal{S}_t^{\epsilon}$ is to its optimum, the more accurate is the certification of the design structure. Notably, we can extend this to account also for the possibility of mixed states by expanding the domain of the frame potential. Define $\tilde{V}_t(\{\rho\})\equiv \sum_{j,k=1}^N F(\rho_{j},\rho_{k})^{2t}$, where $F$ denotes the fidelity.  Since a pair of mixed states cannot be discriminated with success probability larger than $\alpha_\epsilon(1-F)$ when the error rate $\epsilon$ is allowed \cite{Proof}, it follows that also $\tilde{V}_t$ is bounded by the right-hand-side of Eq~\eqref{robust} and thus admits a robust certification.


\textit{Certifying detection efficiency.---}  We turn our attention to the measurement device. A realistic measurement device can be modelled as succeeding with performing the intended detection only with some probability $\eta\in[0,1]$. As a typical example, a single-photon avalanche diode for visibile light has a detection efficiency around $\eta=55\%$ \cite{Detector}. Naturally, it is often practical to infer the efficiency by assuming a simple model for the detector and probing it with single photons. Here, however, we consider the SDI situation in which the overall detection efficiency is bounded based solely on the statistics gathered in the considered experiments.

As before, we appropriately choose $\epsilon$ by inspecting the data $p(b|x,y)$ and accordingly consider the figure of merit \eqref{mod}. In order to certify the detection efficiency, we must consider the optimal value of $\mathcal{S}_t^\epsilon$ that is compatible with a hypothesised value of $\eta$. Since the measurement device is uncharacterised, it can internally map a failed detection onto the outputs $b\in\{1,2,\perp\}$ so that detection failure cannot be read out directly by the experimenter. If failed detections are outputted as $b=1$ or $b=2$, it will sometimes (in half the cases) give a wrong answer to the discrimination task. This sharply increases the error rate $\epsilon$ while making no better contribution to the discrimination than a trivial random guess. Therefore, the device optimally treats failed detections as inconclusive outcomes ($b=\perp$). From Eq~\eqref{haya}, this causes the  success probability in bounded-error discrimination to obey $p_\text{usd}^{\phi_1,\phi_2}\leq \eta\alpha_\epsilon\left(1-|\braket{\phi_{1}}{\phi_{2}}|\right)$ with a possible equality when $\epsilon\leq \epsilon_\text{c}$.

Now, we can evaluate a bound on the optimal quantum value of $\mathcal{S}_t^\epsilon$ when subject to a given amount of detection loss. A simple calculation asserts the following useful inequality 
\begin{equation}\label{lemma}
(\alpha_\epsilon-\max_{M}p_\text{usd}^{\phi_1,\phi_2})^2\geq \alpha_\epsilon^2\left(\left(1-\eta\right)^2+\eta(2-\eta)|\braket{\phi_{1}}{\phi_{2}}|^2\right).
\end{equation}
Applying this inequality for every input $y$, we can bound the figure of merit as follows:
\begin{align}\nonumber\label{errorfree}
& \mathcal{S}_t^\epsilon \geq	 \min_{\{\psi\}}\alpha_\epsilon^{2t} \sum_{y_1<y_2} \left(\left(1-\eta\right)^2+\eta(2-\eta)|\braket{\psi_{y_1}}{\psi_{y_2}}|^2\right)^t=\frac{\alpha_\epsilon}{2}\\\nonumber
&\times\Bigg(-N+\min_{\{\psi\}}\sum_{n=0}^{t}\binom{t}{n}\left(1-\eta\right)^{2(t-n)}\eta^n(2-\eta)^{n}V_{n}(\{\psi\})\Bigg)\\
&\geq \alpha_\epsilon^{2t}\left( -\frac{N}{2}+\frac{1}{2}\sum_{n=0}^{t}\binom{t}{n}\left(1-\eta\right)^{2(t-n)}\eta^n(2-\eta)^{n}J_{n}\right).
\end{align}
In the second line we have used the binomial theorem and identified the $n$'th order frame potential, and in the third line we have used Eq~\eqref{design}. For simplicity, we write $\mathcal{S}_t^\epsilon\geq \alpha_\epsilon^{2t}\mathcal{Q}_t^\eta$ where $\mathcal{Q}_t^\eta$ denotes the bracket on the last line. Hence, for any observed $\mathcal{S}_t^{\epsilon}$, the following $2t$-degree polynomial must be positive: $P(\eta)\equiv \mathcal{S}_{t}^{\epsilon}-\alpha_\epsilon^{2t}\mathcal{Q}_t^\eta\geq 0$. To determine a lower bound on the detection efficiency, we must decide the values of $\eta$ that respect the positivity of $P(\eta)$. This is achieved by finding the real-valued roots (in the interval $[0,1]$) of $P(\eta)$.

\textit{Application: detection efficiency based on SICs.---} While the bound on $\eta$ is typically not tight due to \eqref{lemma}, it enables useful certification. We exemplify this through the previously considered one-parameter family of designs known as SICs (corresponding to $N=d^2$ with $t=2$ for any $d\geq 2$). For this family, we evaluate the relevant root of $P(\eta)$ and find that

\begin{equation}\label{example}
\eta\geq\frac{\alpha_\epsilon d(d-1)-\sqrt{d-1}\sqrt{\sqrt{2} \sqrt{\left(d^2-1\right) \mathcal{S}_2^\epsilon}-\alpha_\epsilon^2d(d-1)}}{\alpha_\epsilon d(d-1)}
\end{equation}
Notice that an optimal implementation (possible when $\epsilon\leq \frac{1}{2}\left(1-\sqrt{\frac{d}{d+1}}\right)$) leads to $\mathcal{S}_2^\epsilon=\alpha_\epsilon^4\frac{d^2(d-1)}{2(d+1)}$ which inserted into Eq~\eqref{example} implies perfect detection efficiency ($\eta=1$). In the other extreme, the bound only becomes trivial ($\eta \geq 0$) when $\mathcal{S}^\epsilon_{2}=\frac{\alpha^4_\epsilon}{2}(d^4-d^2)$, where the second factor is the algebraically maximal (trivial) value of the frame potential. For any intermediate value of $\mathcal{S}_2^\epsilon$, we obtain a non-trivial bound on $\eta$. Notably, this stands in contrast to certification of detection efficiency based on Bell inequality violations for which there exists a (often quite high \cite{Brunner2008}) threshold value for $\eta$ below which no device-independent certification of detection efficiency can be made.

We exemplify the certification in a concrete implementation with imperfections. Take the qubit case of $d=2$ ($N=4$ and $t=2$) and consider that the ideal preparations of Alice and the ideal measurements of Bob are subject to some noise rate $\gamma$. For simplicity, let us concentrate all the noise in Bob's measurements: the optimal measurements for USD are only implemented with probability $1-\gamma$ whereas with probability $\gamma$ the measurement corresponds to a random guess $b\in\{1,2\}$. In addition, we let Bob's device have a detection efficiency of $\eta_\text{exp}=55\%$ and let it treat failed outcomes as $b=\perp$. The combination of noise and detection loss leads to both errors (with probability $\gamma/2$) and sub-optimal correlations corresponding to  $p_\text{usd}^y=\eta_\text{exp}\left((1-\gamma)\left(1-|\braket{\psi_{y_1}}{\psi_{y_2}}|\right)+\gamma/2\right)$. Using that Alice's tetrahedral states have $|\braket{\psi_{y_1}}{\psi_{y_2}}|^2=1/3$, we can establish the error $\epsilon$  through Eq~\eqref{boundederror} and evaluate the certified detection efficiency through Eq~\eqref{example}. For a nearly noise-free implemenetation ($\gamma=0.5\%$) we certify $\eta \geq 31.8\%$. For an order of magnitude higher noise rate ($\gamma=5\%$) we can still certify a detection efficiency of $\eta\geq 21.0\%$.

\textit{Certification of non-projective measurements.---} An interesting feature of USD is that the optimal implementation uses non-projective measurements. Due to the increasing interest in non-projective measurements for quantum information applications, it is relevant to certify such measurement in SDI scenarios. In Supplementary Material we exemplify this for $(N,d,t)=(4,2,2)$ and show that a certification can be achieved for a detection efficiency of at least $\eta= \frac{3+\sqrt{3}}{6}\approx 78.9\%$. This threshold is notable since it is much lower than the nearly perfect detection efficiency required in other SDI schemes based on classically correlated quantum devices \cite{TavakoliSIC, TavakoliNonProj, Smania, Piotr, Lima}


\textit{Conclusions.---} I have developed methods for the certification of state preparation devices and measurement devices in prepare-and-measure experiments in which the devices are assumed to be independent.  The presented scheme is versatile as it applies to three qualitatively different problems: i) certification of quantum states, ii) certification of detection efficiency and iii) certification of non-projective measurements. The certification is robust to errors and therefore applicable to experiments. Notably, small experimental deviations from the assumptions in the SDI scenario, such as memory effects in the detector or multiphoton events, can be accounted for using the method of Ref.~\cite{Lunghi}.

The framework based on independent quantum devices departs from the more common setting in which devices can be classically correlated. This is often natural when considering tasks that are not of adversarial nature. It is therefore relevant to develop such certification schemes targeting various useful properties of quantum systems.  More generally, the loss of convexity that comes with the independence assumption makes it challenging to determine the limitations of quantum correlations and consequently also their applications towards various certification tasks. Here, our tool for overcoming this obstacle relied significantly on USD and the theory of quantum designs. It is of general interest to develop tools for characterising the set of quantum correlations without shared randomness. This would be both of foundational interest and a route to interesting protocols for quantum information processing.

\textit{Note added.---} During the completion of this work, I became aware of the related work of Ref~\cite{Nikolai}.

\acknowledgements
I thank Nicolas Brunner and Jonatan Bohr Brask for discussions and comments. This work was supported by the Swiss National Science Foundation (Starting grant DIAQ, NCCR-QSIT).

\appendix

\section{Randomised USD with stochastic projective measurements}\label{AppProj}
In order to bound the quantum performance of randomised USD under projective measurements, we must first remind ourselves of how projective measurements perform in standard USD. It is a well-known result that for any two pre-established equiprobable non-orthogonal pure states, the optimal USD under stochastic projective measurements is obtained by randomly measuring either the eigenbasis of the first state or the eigenbasis of the second state. This leads to
\begin{equation}
\max_\text{projective} p_\text{usd}=\frac{1-|\braket{\phi_1}{\phi_2}|^2}{2}.
\end{equation}
Naturally, in the special case of orthogonal states, the USD has a unit success rate.

We apply this to randomised USD. For every input $y$, the best success rate reads
\begin{equation}
\max_{\text{projective}} p_{\text{usd}}^y=\frac{1}{2}\left(1-|\braket{\psi_{y_1}}{\psi_{y_2}}|^2+\tau_{\psi_{y_1},\psi_{y_2}}\right),
\end{equation}
where the special case of orthogonal state is accounted for by defining $\tau=1$ if and only if $\braket{\psi_{y_1}}{\psi_{y_2}}=0$ and otherwise $\tau=0$. For any given set of preparations, we can now evaluate the best performance in randomised USD for stochastic projective measurements to be 
\begin{align}\nonumber\label{projsum}
& \mathcal{S}_{t}(\{\psi\})= \frac{1}{2^{2t}}\sum_{y_1<y_2}\left(1+|\braket{\psi_{y_1}}{\psi_{y_2}}|^2-\tau_{\psi_{y_1},\psi_{y_2}}\right)^{2t}\\\nonumber
& =\frac{1}{2^{2t}}\sum_{y_1<y_2}\sum_{n=0}^{2t}\binom{2t}{n}|\braket{\psi_{y_1}}{\psi_{y_2}}|^{2n}\left(1-\tau_{\psi_y,\psi_{y'}}\right)^{2t-n}\\
&= \frac{1}{2^{2t+1}}\sum_{n=1}^{2t}\binom{2t}{n}\left(V_n-N\right)+\frac{1}{2^{2t}}\sum_{y_1<y_2}\left(1-\tau_{\psi_y,\psi_{y'}}\right).
\end{align}
In order to obtain a bound valid for all projective measurements and all state ensembles, we must find a lower bound on the above expression valid for all states. However, this appears not to be straightforward due to the right-most term. 

However, by focusing on the most relevant case of qubits we can solve the problem. Consider the example of  $(N,d,t)=(4,2,2)$.  Since we only have four states, the number of possible pairwise orthogonalities is small. One can exhaustively  consider the different orthogonality configurations that influence the right-most term in \eqref{projsum}. This straightforwardly leads to the finding that the optimal configuration features no orthogonalities among the four states. We can then obtain a lower bound on \eqref{projsum} via the global lower bound on the frame potential (it gives $\mathcal{S}_2\geq 11/10$). However, this bound is sub-optimal since four qubit states cannot be used to form the $3$- or $4$-design that appear in the final expression in \eqref{projsum}. A better bound can be obtained by directly exploiting the Bloch sphere parameterisation to reliably minimise the final expression in \eqref{projsum}. This leads to an optimal configuration being four Bloch vectors pointing to the vertices of a tetrahedron (quantum $2$-design). Since this means $|\braket{\psi_{y_1}}{\psi_{y_2}}|^2=1/3$ for $y_1\neq y_2$. Inserted into Eq~\eqref{projsum}, we obtain that projective measurements must obey $\mathcal{S}_2\geq 32/27$.

The sizable gap between $\mathcal{S}_2\geq 32/27$ and the best quantum result at $\mathcal{S}_2=2/3$ allows for the certification of non-projectiveness to be robust to errors. Consider for instance that Bob's detectors succeed with probability $\eta$. Due to the unambiguity of the discimniation, failed events must be mapped to $b=\perp$. We therefore have that $p_\text{usd}^y=\eta\left(1-|\braket{\psi_{y_1}}{\psi_{y_2}}|\right)=\eta\frac{\sqrt{3}-1}{\sqrt{3}}$. We therefore have
\begin{equation}
\mathcal{S}_t=6\times \left[1-\eta\frac{\sqrt{3}-1}{\sqrt{3}}\right]^4.
\end{equation}
The critical value of $\eta$ for certifying the implementation of non-projective measurements is obtained from solving $\mathcal{S}_t=32/27$. The critical detection efficiency becomes
\begin{equation}
\eta=\frac{3+\sqrt{3}}{6}\approx 78.9\%.
\end{equation}


\begin{thebibliography}{99}
	
	
\bibitem{Vogel}	
K. Vogel and H. Risken,
Determination of quasiprobability distributions in terms of probability distributions for the rotated quadrature phase,
\href{https://doi.org/10.1103/PhysRevA.40.2847}{Phys. Rev. A \textbf{40}, 2847(R) (1989).}

	
\bibitem{Luis}
A. Luis and L. L. S\'anchez-Soto,
Complete Characterization of Arbitrary Quantum Measurement Processes,
\href{https://doi.org/10.1103/PhysRevLett.83.3573}{Phys. Rev. Lett. \textbf{83}, 3573 (1999).}

\bibitem{Rosset}
D. Rosset, R. Ferretti-Sch\"obitz, J-D. Bancal, N. Gisin and Y-C. Liang,
Imperfect measurement settings: Implications for quantum state tomography and entanglement witnesses,
\href{https://doi.org/10.1103/PhysRevA.86.062325}{Phys. Rev. A \textbf{86}, 062325 (2012).}

\bibitem{Anwer}
H. Anwer, S. Muhammad, W. Cherifi, N. Miklin, A. Tavakoli and M. Bourennane,
Experimental characterisation of unsharp qubit measurements in a semi-device-independent setting
\href{https://doi.org/10.1103/PhysRevLett.125.080403}{Phys. Rev. Lett. \textbf{125}, 080403 (2020).}


\bibitem{Tavakoli18}
A. Tavakoli, J. Kaniewski, T. V\'ertesi, D. Rosset, and N. Brunner,
Self-testing quantum states and measurements in the prepare-and-measure scenario,
\href{https://doi.org/10.1103/PhysRevA.98.062307}{Phys. Rev. A \textbf{98}, 062307 (2018).}




\bibitem{Hendrych}
M. Hendrych, R. Gallego, M. Micuda, N. Brunner, A. Acin, and J. P. Torres,
Experimental estimation of the dimension of classical and quantum systems,
\href{https://doi.org/10.1038/nphys2334}{Nature Physics {\bf 8}, 588 (2012).}

\bibitem{Ahrens}
J. Ahrens, P. Badziag, A. Cabello, and M. Bourennane,
Experimental device-independent tests of classical and quantum dimensions,
\href{https://doi.org/10.1038/nphys233}{Nature Physics {\bf 8}, 592 (2012).}

\bibitem{TH15}
A. Tavakoli, A. Hameedi, B. Marques, and M. Bourennane,
Quantum random access codes using single d-level systems,
\href{https://doi.org/10.1103/PhysRevLett.114.170502}{Phys. Rev. Lett. \textbf{114}, 170502 (2015).}

\bibitem{Ambrosio}
V. D'Ambrosio, F. Bisesto, F. Sciarrino, J. F. Barra, G. Lima, and A. Cabello,
Device-independent certification of high-dimensional quantum systems,
\href{https://doi.org/10.1103/PhysRevLett.112.140503}{Phys. Rev. Lett. \textbf{112}, 140503 (2014).}


\bibitem{Lunghi}
T. Lunghi, J. B. Brask, C. C. W. Lim, Q. Lavigne, J. Bowles, A. Martin, H. Zbinden, and N. Brunner,
Self-testing quantum random number generator,
\href{https://doi.org/10.1103/PhysRevLett.114.150501}{Phys. Rev. Lett. {\bf 114}, 150501 (2015).}


\bibitem{Martinez}
D. Mart\'inez, A. Tavakoli, M. Casanova, G. Ca\~nas, B. Marques, and G. Lima,
High-dimensional quantum communication complexity beyond strategies based on Bell's theorem,
\href{https://doi.org/10.1103/PhysRevLett.121.150504}{Phys. Rev. Lett. \textbf{121}, 150504  (2018).}



\bibitem{Farkas19}
M. Farkas and J. Kaniewski,
Self-testing mutually unbiased bases in the prepare-and-measure scenario,
\href{https://doi.org/10.1103/PhysRevA.99.032316}{Phys. Rev. A \textbf{99}, 032316 (2019).}

\bibitem{TavakoliSIC}
A. Tavakoli, D. Rosset, and M-O. Renou,
Enabling Computation of Correlation Bounds for Finite-Dimensional Quantum Systems via Symmetrization,
\href{https://doi.org/10.1103/PhysRevLett.122.070501}{Phys. Rev. Lett. \textbf{122}, 070501 (2019)}


\bibitem{TavakoliNonProj}
A. Tavakoli, M. Smania, T. V\'ertesi, N. Brunner, and M. Bourennane,
Self-testing non-projective quantum measurements in prepare-and-measure experiments,
\href{https://doi.org/10.1126/sciadv.aaw6664}{Science Advances \textbf{6}, 16 (2020).}


\bibitem{Piotr}
P. Mironowicz and M. Paw\l{}owski,
Experimentally feasible semi-device-independent certification of 4 outcome POVMs,
\href{https://doi.org/10.1103/PhysRevA.100.030301}{Phys. Rev. A \textbf{100}, 030301 (2019)}



\bibitem{Mohan}
K. Mohan, A. Tavakoli and N. Brunner,
Sequential random access codes and self-testing of quantum measurement instruments,
\href{https://doi.org/10.1088/1367-2630/ab3773}{New J. Phys. \textbf{21} 083034 (2019).}


\bibitem{Miklin}
N. Miklin, J. J. Borka\l{}a, and M. Paw\l{}owski,
Self-testing of unsharp measurements,
\href{https://doi.org/10.1103/PhysRevResearch.2.033014}{Phys. Rev. Research \textbf{2}, 033014 (2020).}



\bibitem{Smania}
M. Smania, P. Mironowicz, M. Nawareg, M. Paw\l{}owski, A. Cabello, and M. Bourennane,
\href{https://doi.org/10.1364/OPTICA.377959}{Optica \textbf{7}, 123 (2020).}

\bibitem{Foletto}
G. Foletto, L. Calderaro, G. Vallone and P. Villoresi,
Experimental demonstration of sequential quantum random access codes,
\href{https://doi.org/10.1103/PhysRevResearch.2.033205}{Phys. Rev. Research \textbf{2}, 033205  (2020).}



\bibitem{Bowles}	
J. Bowles, M. T. Quintino and N. Brunner,
Certifying the dimension of classical and quantum systems in a prepare-and-measure scenario with independent devices,
\href{https://doi.org/10.1103/PhysRevLett.112.140407}{Phys. Rev. Lett. \textbf{112}, 140407 (2014).}



\bibitem{Vicente}	
J. I. de Vicente,
Shared randomness and device-independent dimension witnessing,
\href{https://doi.org/10.1103/PhysRevA.95.012340}{Phys. Rev. A \textbf{95}, 012340 (2017).}
	
	
	


\bibitem{Ivanovic}
I. D. Ivanovic,
How to differentiate between non-orthogonal states,
\href{https://doi.org/10.1016/0375-9601(87)90222-2}{Phys. Lett. A \textbf{123}, 257 (1987).}

\bibitem{Dieks}
D. Dieks,
Overlap and distinguishability of quantum states,
\href{https://doi.org/10.1016/0375-9601(88)90840-7}{Phys. Lett. A \textbf{126}, 303 (1988).}

\bibitem{Peres}
A. Peres,
How to differentiate between non-orthogonal states,
\href{https://doi.org/10.1016/0375-9601(88)91034-1}{ Phys. Lett. A \textbf{128}, 19 (1988).}


\bibitem{Delsarte}
P. Delsarte, J. M. Goethals and J. J. Seidel,
Spherical codes and designs,
\href{https://doi.org/10.1007/BF03187604}{Geom. Dedicata \textbf{6}, 363 (1977).}


\bibitem{Renes}
J. M. Renes, R. Blume-Kohout, A. J. Scott, C. M. Caves,
Symmetric Informationally Complete Quantum Measurements,
\href{https://doi.org/10.1063/1.1737053}{J. Math. Phys. \textbf{45}, 2171 (2004).}



\bibitem{Hayashi}
A. Hayashi, T. Hashimoto and M. Horibe,
Reexamination of optimal quantum state estimation of pure states,
\href{https://doi.org/10.1103/PhysRevA.72.032325}{Phys. Rev. A \textbf{72}, 032325 (2005).}

\bibitem{Scott}
A. Scott, 
Optimizing quantum process tomography with unitary 2-designs,
\href{https://doi.org/10.1088/1751-8113/41/5/055308}{ J. Phys. A \textbf{41}, 055308 (2008).}


\bibitem{SixState}
D. Bru\ss,
Optimal Eavesdropping in Quantum Cryptography with Six States,
\href{https://doi.org/10.1103/PhysRevLett.81.3018}{Phys. Rev. Lett. \textbf{81}, 3018 (1998).}

\bibitem{Joe}
J. M. Renes,
Spherical-code key-distribution protocols for qubits,
\href{https://doi.org/10.1103/PhysRevA.70.052314}{Phys. Rev. A \textbf{70}, 052314 (2004).}

\bibitem{MUBSIC}
A. Tavakoli, M. Farkas, D. Rosset, J-D. Bancal and J. Kaniewski,
Mutually unbiased bases and symmetric informationally complete measurements in Bell experiments: Bell inequalities, device-independent certification and applications,
\href{https://arxiv.org/abs/1912.03225}{arXiv:1912.03225}



\bibitem{Ketterer}
A. Ketterer and O. G\"uhne,
Entropic uncertainty relations from quantum designs,
\href{https://arxiv.org/abs/1911.07533}{arXiv:1911.07533}

\bibitem{Lloyd}
Z-W. Liu, S. Lloyd, E. Y. Zhu, and H. Zhu,
Generalized Entanglement Entropies of Quantum Designs,
\href{https://doi.org/10.1103/PhysRevLett.120.130502}{Phys. Rev. Lett. \textbf{120}, 130502 (2018).}

\bibitem{Bae}
J. Bae, B. C Hiesmayr and D. McNulty,
Linking entanglement detection and state tomography via quantum 2-designs,
\href{https://doi.org/10.1088/1367-2630/aaf8cf}{New J. Phys. \textbf{21} 013012 (2019).}








\bibitem{MUBs}
T. Durt, B-G. Englert, I. Bengtsson and K. \.Zyczkowski,
On mutually unbiased bases,
\href{https://doi.org/10.1142/S0219749910006502}{Int. J. Quantum Information \text{8}, 535 (2010).}

\bibitem{SICs}
C. A. Fuchs, M. C. Hoang, and B. C. Stacey,
The SIC Question: History and State of Play,
\href{https://doi.org/10.3390/axioms6030021}{Axioms \textbf{21}, 6 (2017).}
	
\bibitem{Plato}
A. Tavakoli and N. Gisin,
The Platonic solids and fundamental tests of quantum mechanics,
\href{https://arxiv.org/abs/2001.00188}{arXiv:2001.00188}
	
	
\bibitem{Fickus}
J. J. Benedetto and M. Fickus,
Finite Normalized Tight Frames,
\href{https://doi.org/10.1023/A:1021323312367}{Adv. Comput. Math. \textbf{18}, 357 (2003).}
	
	

	
\bibitem{Hayashi2}
A. Hayashi, T. Hashimoto, and M. Horibe,
State discrimination with error margin and its locality,
\href{https://doi.org/10.1103/PhysRevA.78.012333}{Phys. Rev. A \textbf{78}, 012333 (2008).}

\bibitem{Clarke}
R. B. M. Clarke, A. Chefles, S. M. Barnett and E. Riis,
Experimental demonstration of optimal unambiguous state discrimination,
\href{https://doi.org/10.1103/PhysRevA.63.040305}{Phys. Rev. A \textbf{63}, 040305(R) (2001).}

\bibitem{Proof}
This is straightforwardly proven by first purifying the two mixed states, exploiting the known results for error-bounded discrimination of pure states \cite{Hayashi2} and lastly using Uhlmann's theorem.

\bibitem{Detector}
F. Villa, D. Bronzi, Y. Zou, C. Scarcella, G. Boso, S. Tisa, A. Tosi, F. Zappa, D. Durini, S. Weyers, U. Paschen and W. Brockherde,
CMOS SPADs with up to 500 $\mu$m diameter and 55\% detection efficiency at 420 nm,
\href{https://doi.org/10.1080/09500340.2013.864425}{Journal of Modern Optics \textbf{61}, 102 (2014).}		

\bibitem{Brunner2008}
N. Brunner and N. Gisin,
Partial list of bipartite Bell inequalities with four binary settings,
\href{https://doi.org/10.1016/j.physleta.2008.01.052}{Phys. Lett. A \textbf{372},  3162 (2008).}
	
\bibitem{Lima}
E. S. G\'omez, S. G\'omez, P. Gonz\'alez, G. Ca\~nas, J. F. Barra, A. Delgado, G. B. Xavier, A. Cabello, M. Kleinmann, T. V\'ertesi, and G. Lima,
Device-Independent Certification of a Nonprojective Qubit Measurement,
\href{https://doi.org/10.1103/PhysRevLett.117.260401}{	Phys. Rev. Lett. \textbf{117}, 260401 (2016)}	
	
\bibitem{Nikolai}
N. Miklin and M. Oszmaniec,
A universal scheme for robust self-testing in the prepare-and-measure scenario,
arXiv:2003.01032 (03-03-2020).	


\end{thebibliography}
\end{document}